\documentclass[submit]{smj}

\makeatletter
\def\maxwidth{ %
  \ifdim\Gin@nat@width>\linewidth
    \linewidth
  \else
    \Gin@nat@width
  \fi
}
\makeatother

\definecolor{fgcolor}{rgb}{0.345, 0.345, 0.345}
\newcommand{\hlnum}[1]{\textcolor[rgb]{0.686,0.059,0.569}{#1}}%
\newcommand{\hlstr}[1]{\textcolor[rgb]{0.192,0.494,0.8}{#1}}%
\newcommand{\hlopt}[1]{\textcolor[rgb]{0,0,0}{#1}}%
\newcommand{\hlstd}[1]{\textcolor[rgb]{0.345,0.345,0.345}{#1}}%
\newcommand{\hlkwb}[1]{\textcolor[rgb]{0.69,0.353,0.396}{#1}}%
\newcommand{\hlkwc}[1]{\textcolor[rgb]{0.333,0.667,0.333}{#1}}%
\newcommand{\hlkwd}[1]{\textcolor[rgb]{0.737,0.353,0.396}{\textbf{#1}}}%

\usepackage{framed}
\makeatletter
\newenvironment{kframe}{%
 \def\at@end@of@kframe{}%
 \ifinner\ifhmode%
  \def\at@end@of@kframe{\end{minipage}}%
  \begin{minipage}{\columnwidth}%
 \fi\fi%
 \def\FrameCommand##1{\hskip\@totalleftmargin \hskip-\fboxsep
 \colorbox{shadecolor}{##1}\hskip-\fboxsep
     \hskip-\linewidth \hskip-\@totalleftmargin \hskip\columnwidth}%
 \MakeFramed {\advance\hsize-\width
   \@totalleftmargin\z@ \linewidth\hsize
   \@setminipage}}%
 {\par\unskip\endMakeFramed%
 \at@end@of@kframe}
\makeatother

\definecolor{shadecolor}{rgb}{.97, .97, .97}
\definecolor{messagecolor}{rgb}{0, 0, 0}
\definecolor{warningcolor}{rgb}{1, 0, 1}
\definecolor{errorcolor}{rgb}{1, 0, 0}
\newenvironment{knitrout}{}{} 

\usepackage{alltt}
\usepackage{float}
\usepackage{natbib}
\usepackage[english]{babel}
\usepackage{tabularx}
\usepackage{lscape}
\usepackage{multirow}
\usepackage{rotating}
\usepackage{dsfont}
\usepackage{subfig}
\usepackage{bbm}
\usepackage{wrapfig}
\usepackage{booktabs}
\usepackage{enumitem}

\usepackage{calc}
\usepackage{ifthen}
\newenvironment{tabularsmall}
{ \footnotesize \sffamily \tabular } {
\endtabular
\normalfont }

\newcommand{\gammab}{\boldsymbol{\gamma}}

\newcommand{\xb}{\boldsymbol{x}}

\newcommand{\0}{\textbf{0}}

\newcommand{\blanco}[1]{}

\usepackage{natbib}

\Author{Moritz Berger\Affil{1} and Matthias Schmid\Affil{1}}
\AuthorRunning{Moritz Berger and Matthias Schmid}
  
\Affiliations{

\item Department of Medical Biometry, Informatics and Epidemiology, 
      Faculty of Medicine,
      University of Bonn, 
			Germany 
}

\CorrAddress{Moritz Berger, 
             Department of Medical Biometry, Informatics and Epidemiology, 
             Faculty of Medicine,
             University of Bonn, 
             Siegmund-Freud-Stra{\ss}e 25, 
             53127 Bonn, 
             Germany}
\CorrEmail{Moritz.Berger@imbie.uni-bonn.de}
\CorrPhone{(+49)\;228\;287\;14096}
\CorrFax{(+49)\;228\;287\;11324}

\Title{Semiparametric Regression for Discrete Time-to-Event Data}
\TitleRunning{Discrete Time-to-Event Models}

\Abstract{
Time-to-event models are a popular tool to analyse data where the outcome variable is
the time to the occurrence of a specific event of interest. Here we focus on the analysis of time-to-event outcomes that are either intrisically discrete or grouped versions of continuous event times. In the literature, there exists a variety of regression methods for such data. This tutorial provides an introduction to how these models can be applied using open source statistical software. In particular, we consider semiparametric extensions comprising the use of smooth nonlinear functions and tree-based methods. All methods are illustrated by data on the duration of unemployment of U.S.\@ citizens.
}


\begin{document}

\maketitle

\section{Introduction}

The objective of many statistical analyses is to model a duration time until a specifc event occurs. This is usually referred to as {\em time-to-event} or {\em survival analysis}. In biostatistics, for example, one often examines the time to death or the progression of a disease. In economics and the social sciences, popular examples include the modeling of the duration of unemployment or the time to retirement. Generally, in regression models for time-to-event data the event time itself is the response variable, and one wants to investigate the association of the response with several explanatory variables. Most often it is assumed in these analyses that the survival time is given by a random variable measured on a {\em continuous scale}. This case has been studied extensively in the literature, see, for example, \citet{Kalbfleisch2002} and \citet{Klein2003}. However, in practice measurements of time are often discrete. Durations, for example, are often measured in days, years or months. Moreover, there are situations where the exact event time may not be known, but only an interval during which the event of interest took place. 

Here we consider the application of regression models for {\em discrete} time-to-event data, which are characterized by an ordinal response variable taking the numbers $1,2, \hdots$ . These numbers either refer to a situation where event times are intrinsically discrete (such as the time to pregnancy, which in clinical applications is usually measured by the number of menstrual cycles), or when continuous event times have been grouped. In the latter case, the numbers $t = 1, 2,\hdots$ , refer to mutually exclusive time intervals $[0, a_1)$, $[a_1, a_2)$, $\hdots$ , with fixed boundaries $a_1, a_2, \hdots$ .
Generally, a great advantage of discrete time-to-event models is that they can be viewed as regression models with binary response, giving rise, e.g., to the application of logistic regression or probit regression \citep{willettsinger93}.

A comprehensive treatment of the statistical methodology for discrete time-to-event data has recently been given by \citet{Tutz2016}. Similar to Gaussian regression, a large part of this methodology has been designed to estimate predictor-response relationships using a {\em linear} combination of the exaplanatory variables. In addition, \citet{Tutz2016} discuss several (less well known) approaches for {\em semiparametric} discrete time-to-event modeling. The aim of this tutorial is to provide an in-depth explanation of how these semiparametric models can be fitted and implemented using the \texttt{R} software for statistical computing \citep{R2017}. In particular, we will explain how smooth nonlinear functions and tree-based methods can be incorporated into discrete time-to-event models.

A frequently observed phenomenon in time-to-event analysis is censoring. Generally, a duration time is termed ``censored'' if its total length has not been fully observed. In this article we consider the most common type of {\em type-I} or {\em right censoring}, which means that the beginnings of the duration times are observed for all individuals in a study, whereas the respective ends are only observed for part of the individuals. Hence for some of the individuals it is only known that the event occured later than the observed time.

\begin{table}[!t]
\caption{Summary statistics of the six explanatory variables used in the modeling of the U.S.\@ unemployment data ($n=3,210$).}
\begin{center}
\begin{tabularsmall}{lrrrrrr}
\toprule
Variable&\multicolumn{6}{c}{Summary statistics}\\
\midrule
&$x_{min}$&$x_{0.25}$&$x_{med}$&$\bar{x}$&$x_{0.75}$&$x_{max}$\\
\texttt{age}&20&27&34&35.45&43&61\\
\texttt{reprate}&0.06&0.39&0.50&0.45&0.52&2.05\\
\texttt{disrate}&0.01&0.05&0.10&0.11&0.15&1.02\\
\texttt{logwage}&2.70&5.29&5.68&5.69&6.05&7.60\\
\texttt{tenure}&0&0&2&4.11&5&40\\
\\[-.3cm]
\texttt{ui}&\multicolumn{3}{c}{no: 1437 ($44.8\%$)}&\multicolumn{3}{c}{yes: 1773 ($55.2\%$)}\\
\bottomrule
\end{tabularsmall}
\end{center}
\label{tab:data_summary}
\end{table}

\begin{figure}[!t]
\centering
\includegraphics[width=0.6\textwidth]{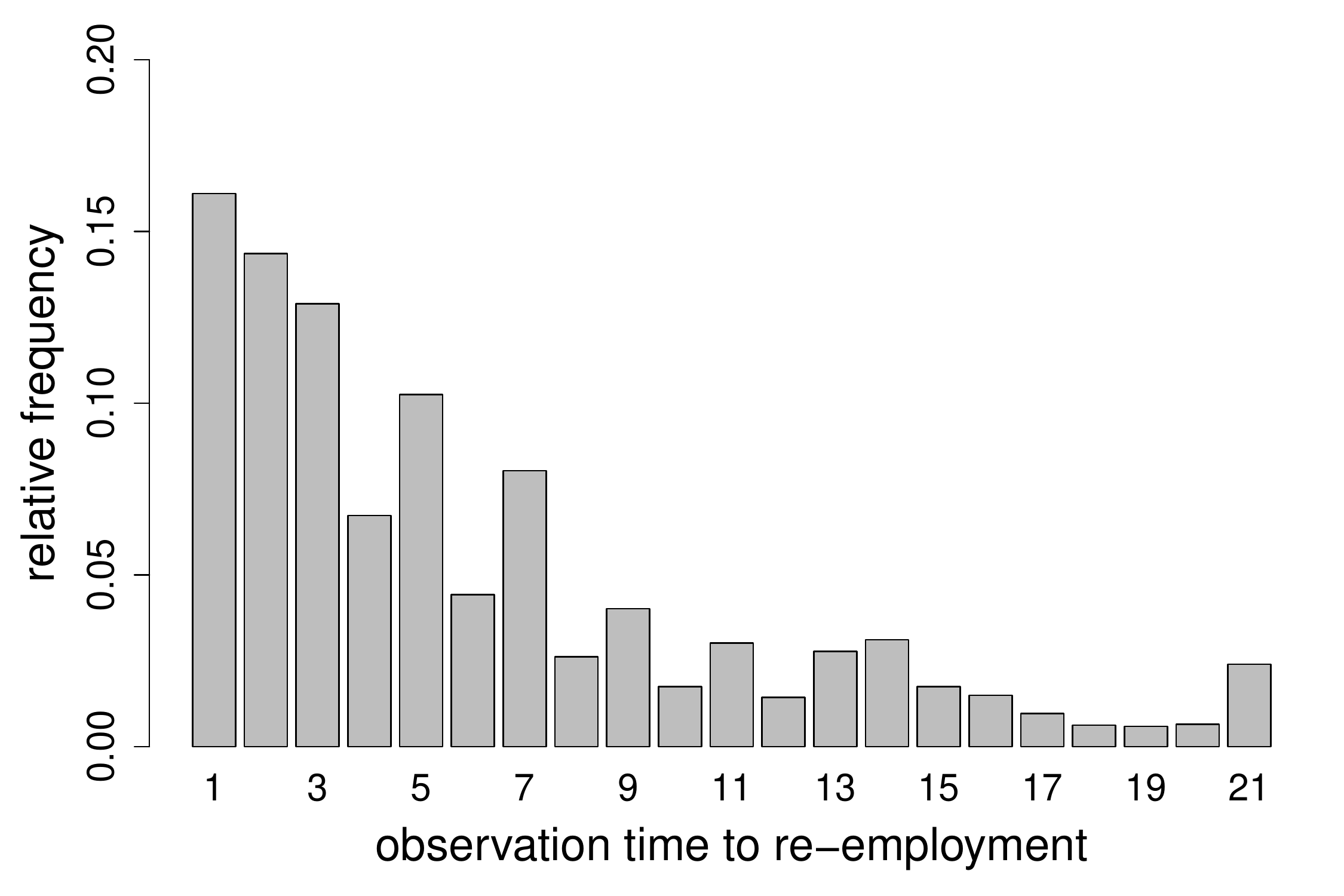}
\caption{Observed time to re-employment (measured in two-week intervals) in the U.S.\@ unemployment data. The median observation time in the data is 4, corresponding to a time period of 8 weeks.}
\label{fig:data_time}
\end{figure}

All models discussed in this article will be illustrated by means of a publicly available data set on the duration of unemployment. The data comprise observations obtained from $n=3,343$ U.S.\@ citizens and were collected between 1986 and 1992 as part of the January Current Population Surveys Displaced Workers Supplements
(DWS). The original data set is available as part of the \texttt{R} add-on package \texttt{Ecdat} \citep{Ecdat2015}. The response variable that will be considered here is the time to re-employment in any kind of job, which includes full-time, part-time or other kind of jobs. Due to the study design, the observed unemployment durations are discrete, as they were measured in two-week intervals. In this article, we will analyze the data over a period of 40 weeks comprising 21 possible event times $t=1, 2, \hdots , 21$, where $t=21$ refers to event times $>40$ weeks. Explanatory variables that will be included in our analyses are the age in years (\texttt{age}), an indicator on whether an unemployment insurance claim was submitted (\texttt{ui}), the eligible replacement rate (\texttt{reprate}, defined by the weekly benefit amount divided by the amount of weekly earnings in the lost job), the eligible disregard rate (\texttt{disrate}, defined as the amount up to which recipients of unemployment insurance who accept part-time work can earn without any reduction in unemployment benefits divided by the weekly earnings in the lost job), the log weekly earnings in
the lost job in \$ (\texttt{logwage}), and the tenure in the lost job in years (\texttt{tenure}). The summary statistics of the six explanatory variables are presented in Table \ref{tab:data_summary}. Due to missing values in the variables some observations were exluded from the data, arriving at a sample containing the data of $3,210$ citizens. 

The observed times to re-employment are visualized in Figure \ref{fig:data_time}. If an individual is still jobless at the end of the survey (i.e., after 40 weeks) or dropped out of the study before finding a job it is subject to right censoring. In this case its observation time corresponds to the {\em censoring time}, otherwise to the true time of re-employment. 

The rest of this tutorial is organized as follows: Section \ref{sec:basic} provides the basic theoretical framework and an introduction to parametric as well as semiparametric discrete time-to-event modeling. Details on model fitting and data preparation are given in Section~\ref{sec:est}. Section \ref{sec:gof} presents measures that are useful for assessing the goodness-of-fit of discrete time-to-event models. In Section \ref{sec:app} we illustrate the methods by presenting a detailed analysis of the U.S.\@ unemployment data, showing how the various regression models can be applied by the use of \texttt{R}. Section \ref{sec:remarks} discusses additional aspects related to discrete time-to-event modeling and puts the methods considered in this article into perspective.

The \texttt{R} code to reproduce all the numerical results is provided as electronic supplement to this tutorial.

\section{Notation and Basic Concepts}\label{sec:basic}

Given $n$ observations, $i=1,\hdots,n$, let in the following $T_i$ denote the event time and $C_i$ the censoring time of individual $i$. $T_i$ and $C_i$ are assumed to be independent random variables taking discrete values in $\{1,\hdots,k\}$. In addition one observes a vector of $p$ explanatory variables $\xb_i=(x_{i1},\hdots,x_{ip})^\top$. For right censored data, the observation time is defined by $\tilde{T}_i=\min(T_i,C_i)$, i.e., $\tilde{T}_i$ corresponds to the true event time if $T_i<C_i$ and to the censoring time otherwise. If originally continuous data have been grouped, the discrete event times $1,\hdots,k$ refer to time intervals $[0,a_1), [a_1, a_2), \hdots, [a_{k-1}, \infty )$, where $T_i=t$ means that the event occured in time interval $[a_{t-1},a_t)$. For example, in our application on unemployment durations, where time was measured in two-week intervals, $T_i = 3$ implies that re-employment of individual $i$ took place between four and six weeks after the start of the study.

The main tool to model discrete time-to-event data is the \textit{hazard function}, which captures the dynamics of the survival process at each time point. For a given vector of explanatory variables $\xb_i$, the hazard function is defined by  
\begin{equation}\label{hazard}
\lambda(t|\xb_i)=P(T_i=t \, | \, T_i\geq t,\xb_i),\quad t=1,\hdots,k, 
\end{equation}
describing the conditional probability of an event at time $t$ given that the individual survived until $t$. The corresponding \textit{survival function} is given by 
\begin{equation}\label{survival}
S(t|\xb_i)=P(T_i>t|\xb_i)=\prod_{s=1}^{t}{\left(1-\lambda(s|\xb_i)\right)},\quad t=1,\hdots,k,
\end{equation}
denoting the probability that an event occurs later than at time $t$, or, alternatively, the probability of surviving interval $[a_{t-1},a_t)$. 

An important consequence of the definition of the hazard function in (\ref{hazard}) is that for a fixed time $t$, the hazard $\lambda(t|\xb_i)$ describes a binary variable that distinguishes between the event taking place at time $t$ or not, conditional on $T_i\geq t$. Therefore, a model for the discrete hazard function can be derived from regression modeling strategies for a binary response data.

\subsection{Parametric Discrete Hazard Models}\label{sec:parametric}

A general class of binary response models applied to the discrete hazard function is defined by 
\begin{equation}\label{model_generic}
\lambda(t|\xb_i) = h(\gamma_{0t}+\xb_i^\top\gammab),
\end{equation}
where $h(\cdot )$ is a strictly monotone increasing distribution function. A common assumption is that the model contains a time-varying intercept and a set of covariate effects that are fixed over time. Hence, the linear predictor of the model, $\eta_{it}=\gamma_{0t}+\xb_i^\top\gammab$, comprises the intercepts $\gamma_{0t}$, $t=1,\ldots , k-1$, and a vector of regression coefficients $\gammab=(\gamma_1,\hdots,\gamma_p)^\top$ independent of $t$. Note that there is no intercept parameter for $t=k$, as the hazard function in (\ref{hazard}) is fully determined by $h(\cdot )$ and the coefficients $\gamma_{01},\ldots , \gamma_{0,k-1}, \gammab^\top$.

The most popular version of model \eqref{model_generic} is the {\it logistic discrete hazard model} or  {\it proportional continuation ratio model}, which is specified by the equation
\begin{equation}\label{model_response}
\lambda(t|\xb_i)=\frac{\exp(\gamma_{0t}+\xb_i^\top\gammab)}{1+\exp(\gamma_{0t}+\xb_i^\top\gammab)}.
\end{equation}
By definition, the proportional continuation ratio model uses the logistic distribution function for  $h(\cdot )$. It can be shown that an alternative representation of the model is
\begin{equation}\label{model_continuation}
\log\left(\frac{P(T_i=t|\xb_i)}{P(T_i>t|\xb_i)}\right)=\gamma_{0t}+\xb_i^\top\gammab.
\end{equation}
The ratio $P(T_i=t|\xb_i)/P(T_i>t|\xb_i)$ compares the probability of an event at time~$t$ to the probability of an event later than $t$. It is also known as {\em continuation ratio}, see, for example, \citet{Agresti2013}. This representation of the model allows for an easy interpretation of the effects, see the application in Section \ref{subsec:models}.

Generally, the number of parameters in model \eqref{model_response} depends on the number of time points, as there is a seperate intercept for each $t$. The set of intercepts $\gamma_{01},\hdots,\gamma_{0,k-1}$ defines the hazard that is always present for any given set of covariates. This hazard is ususally refered to as \textit{baseline hazard}, and the intercepts $\gamma_{0t}$ correspond to the log continuation ratio when all covariates are zero.  

\subsection{Semiparametric Extensions}\label{sec:semi}

The parametric model introduced in the previous section is linear in $\gammab$, implying that each covariate has a linear effect on the transformed hazard. In practice, this linearity assumption may be too restrictive, as predictor-response relationships are often characterized by  nonlinear functional forms. Furthermore, it is assumed that the baseline hazard is represented by a separate intercept coefficient for each $t$. This can lead to numerical problems, especially when the number of time points (and hence the number of intercept parameters) is large relative to the sample size, implying that the event counts at some of these time points may become small. In the following we will consider popular semiparametric alternatives for the definition of $\eta$ that address these issues. We first introduce {\it additive hazard models} and subsequently {\it tree-based methods}, which can also be embeeded in the framework of binary response models. 

To avoid numerical problems in the estimation of the baseline hazard, it is often convenient to consider an additive model with predictor 
\begin{equation}\label{eta_smooth_base}
\eta_{it}=f_{0}(t)+\xb_i^\top\gammab,
\end{equation}
where $f_{0}(t)$ is a smooth (possibly nonlinear) function of time. By relating the values of the baseline hazard at neighboring time points via $f_{0}(t)$, the number of parameters involved in model fitting effectively reduces, and low event counts at some time points become less problematic. A common way to specify the smooth function in $t$ is to use splines, which are represented by a weighted sum of $m$ basis functions. One possible representation of $f_{0}(t)$ is by $B$-spline basis functions. These are polynomials of fixed degree $d$ differing from zero in $d+1$ adjacing intervals. For a comprehensive introduction to $B$-splines, see \citet{DeBoor1978}. Very flexible spline functions can be obtained by choosing a relatively large number of basis functions $m$ and at the same time using a penalty term to prevent estimates becoming too rough (``wiggly''). This approach, on which we will focus in this article, is called \textit{P-splines} and was first proposed by \citet{Eilers1996}.

An extension of the semiparametric model \eqref{eta_smooth_base} that weakens the linearity assumption on the effects of the covariates is given by the additive model
\begin{equation}\label{eta_smooth_cov}
\eta_{it}=f_{0}(t)+\sum_{j=1}^{p}{f_j(x_{ij})},
\end{equation}
where the $f_j(x_j)$ are unknown smooth functions. That is, the effects of the covariates (or subsets of the covariates) are determined by smooth, possibly nonlinear, functions. A common approach is again to use $P$-splines and to expand each function seperately by a weighted sum of $B$-spline basis functions depending on the covariates.

In the semiparametric models with predictors \eqref{eta_smooth_base} and \eqref{eta_smooth_cov} it is assumed that the predictor is given by an additive function of time and a linear (or additive) function of the covariates. Although these models are very flexible, they may not capture the structure of the data very well if interactions between covariates are present. For example, it is quite conceivable that the effect of a covariate on the hazard depends on the values of a second covariate, implying the presence of an interaction between the two covariates. The problem when incorporating interactions in parametric or additive models is that the relevant interactions have to be known and specified before model fitting. Furthermore, parametric and additive models are hard to handle if the interaction terms involve more than two covariates. An alternative regression approach that addresses these problems is {\em recursive partitioning}, which is also known as {\em tree modeling}. The most popular tree method is {\em classification and regression trees} (CART), as proposed and described in detail by \citet{Breiman1984}. The basic CART method is conceptually very simple: The covariate space is partitioned recursively
into a set of rectangles, and in each rectangle a simple model (for example, a
covariate-free model) is fitted. A user-friendly introduction to the basic concepts of tree modeling is found
in \citet{HasTibFri2009}. Recently \citet{Schmid2016} proposed a recursive partitioning method that is specifically designed to model discrete time-to-event data. The main principle is to fit a discrete hazard model of the form
\begin{equation}\label{eta_tree}
\lambda(t|\xb_i) =f(t,\xb_i), 
\end{equation}
where $f(t,\xb)$ is represented by a classification tree with binary outcome. Each split of this tree is determined by either $t$ (treated as an ordinal variable) or one of the covariates. As a result, each terminal node of the tree refers to an estimate of the hazard function for a specific covariate combination and a specific time interval $[t_1, t_2] \subset [1, k]$. For details on the calculation of the estimates, see Section \ref{subsec:tree}.

\section{Estimation and Data Preparation for Additive Hazard Models}\label{sec:est}

To derive the log-likelihood function for discrete hazard models it is useful to introduce a binary variable indicating whether the target event was observed or not: 
\begin{equation}\label{delta}
\Delta_i=\begin{cases}1, \;\text{ if }\; T_i \leq C_i,\\0, \;\text{ if }\; T_i>C_i.\end{cases}
\end{equation}
Thus, $\Delta_i$ becomes 1 if the exact true time is observed; otherwise, $\Delta_i=0$.
In the case where continuous time-to-event data are grouped, $\Delta_i = 1$ and $\tilde{T}_i = t$ implies an event in interval $[a_{t-1},a_t)$ and $T_i = \tilde{T}_i = t$.
Similarly, $\Delta_i = 0$ and $\tilde{T}_i = t$ implies $C_i=\tilde{T}_i = t$ and survival beyond $a_t$, i.e., $T_i > \tilde{T}_i = t$.

Note that when continuous time-to-event data are grouped or rounded, additional assumptions are implicitly imposed on the censoring mechanism. To see this, consider the case where both the continuous event time $T_{\mathrm{cont},i}$ and the continuous censoring time $C_{\mathrm{cont},i}$ are within the same interval, say, $[a_{t-1},a_t)$. Then, by definition, $T_i = C_i = t$, $\tilde{T}_i = t$, and $\Delta_i = 1$, leading to the usual interpretation that an event was observed in interval $[a_{t-1},a_t)$. At the same time, however, this interpretation implicitly assumes $T_{\mathrm{cont},i} \le C_{\mathrm{cont},i}$, i.e., the continuous event time $T_{\mathrm{cont},i} \in [a_{t-1},a_t)$ is not allowed to be larger than the continuous censoring time $C_{\mathrm{cont},i} \in [a_{t-1},a_t)$. Without this assumption, the scenario where both $T_{\mathrm{cont},i}, C_{\mathrm{cont},i} \in [a_{t-1},a_t)$ and $T_{\mathrm{cont},i} > C_{\mathrm{cont},i}$ would result in $\tilde{T}_i = t$ and $\Delta_i = 1$ but {\em no} observed event in $[a_{t-1},a_t)$, implying that the usual interpretation of $\tilde{T}_i = t$, $\Delta_i = 1$ would no longer be appropriate. This assumption on the nature of the censoring mechanism is often referred to as ``censoring at the end of the interval''.

With data $(\tilde{T}_i,\Delta_i,\xb_i),\; i=1,\hdots,n$, the contribution of the $i$-th observation to the likelihood function is given by
\begin{equation}\label{Likelihood_full}
L_i=P(T_i=\tilde{T}_i)^{\Delta_i} \, P(T_i>\tilde{T}_i)^{1-\Delta_i}\, P(C_i\geq \tilde{T}_i)^{\Delta_i}\, P(C_i=\tilde{T}_i)^{1-\Delta_i}. 
\end{equation}
A crucial assumption that is usually made to simplify the likelihood function \eqref{Likelihood_full} is that the censoring process does not depend on the parameters determining the event times $T_i$. A consequence of this assumption is that the terms involving the censoring times can be ignored in the maximization of the likelihood function for the time-to-event process. Omitting the terms involving $C_i$ in (\ref{Likelihood_full}) and inserting the definitions of the hazard function \eqref{hazard} and the survival function \eqref{survival}, one obtains (expect for some constants)
\begin{equation}\label{Likelihood}
L_i \,\propto\, \lambda(\tilde{T}_i|\xb_i)^{\Delta_i}(1-\lambda(\tilde{T}_i|\xb_i))^{1-\Delta_i}\prod_{j=1}^{\tilde{T}_i-1}(1-\lambda(j|\xb_i)).
\end{equation}
Note that, by definition, one always obtains $\Delta_i = 1$ and $\lambda(\tilde{T}_i|\xb_i) = 1$ if $\tilde{T}_i$ is equal to the last time point $k$. For maximum likelihood estimation, it is therefore convenient to re-code observations with $\tilde{T}_i = k$ as follows:
\begin{equation}
\tilde{T}_i = k,\,\Delta_i = 1,\,\xb_i \ \, \longmapsto \ \,
\tilde{T}_i = k-1,\,\Delta_i = 0 ,\,\xb_i \ ,
\end{equation}
making use of the fact that the value of the likelihood contribution in (\ref{Likelihood}) will not be altered by this transformation.

With some algebra it can be shown that the likelihood function \eqref{Likelihood} is equal to the likelihood of a binary response model with outcome variables
\begin{equation}\label{responses}
(y_{i1},\hdots,y_{i\tilde{T}_i})=\begin{cases}(0,\hdots,0,1), \;\text{ if }\; \Delta_i=1\\(0,\hdots,0,0), \;\text{ if }\; \Delta_i=0.\end{cases}
\end{equation}
For individuals where the exact event time is observed one defines the observation vector $(0,\hdots,0,1)$ of length $\tilde{T}_i$. For censored individuals the observation vector contains only zeros. According to this definition one has $\tilde{T}_i$ binary observations for each individual $i$, resulting in a total of $\tilde{T}_1+ \hdots +\tilde{T}_n$ observations. Using these definitions, the log-likelihood of the proportional continuation ratio model becomes
\begin{equation}\label{logL}
l \, \propto \, \sum_{i=1}^{n}\sum_{s=1}^{\tilde{T}_i}{y_{is} \log(\lambda(s|\xb_i))+(1-y_{is}) \log(1-\lambda(s|\xb_i))}.
\end{equation}
The main advantage of this representation of the log-likelihood is that it allows to use software for fitting binary response models. For example, it follows from (\ref{logL}) that fitting a continuation ratio model is equivalent to fitting a logistic regression model with predictor \eqref{eta_smooth_base} or \eqref{eta_smooth_cov}. In this model the values of the binary responses $y_{is}$ can be interpreted as binary decisions for the transition from interval $[a_{s-1},a_s)$ to $[a_s,a_{s+1})$. For instance, in the application on unemployment duration one observes $y_{is}=0$ for each two-week interval as long as the individual $i$ is not re-employed yet.

The models with smooth components \eqref{eta_smooth_base} and \eqref{eta_smooth_cov} can be fitted by maximizing a penalized likelihood of the form 
\begin{equation} \label{pen_ll}
\ell_p = \ell - \delta J,
\end{equation}
where $\delta\in \mathbb{R}^+$ is a penalty parameter and $J \in\mathbb{R}^+$ is the penalty term mentioned in Section \ref{sec:semi} putting restrictions on the weights of the $B$-spline basis functions and preventing estimates from becoming too rough. When using $P$-splines, $J$ is a difference penalty on adjacent $B$-spline coefficients, see, e.g., \citet{Eilers1996} for details. A common procedure is to use cubic B-splines ($d=3$) with second order differences. The degree of smoothness is determined by the tuning parameter $\delta$. The larger the value of $\delta$, the smoother is the resulting function, and vice versa. When several smooth functions are included in the model, one uses a difference penalty for each spline effect, based on the differences of adjacent $B$-spline coefficients for the corresponding covariate. The smoothness of the individual spline estimates is determined by seperate penalty parameters $\delta_j$. 

Before fitting proportional continuation ratio models with software for binary outcome data, one has to generate the required binary observations presented in \eqref{responses}. This is done by the generation of an {\em augmented data matrix}. For the setup of the matrix one has to distinguish between censored and non-censored individuals. For an individual whose event was observed ($\Delta_i=1$) at time $\tilde{T}_i$ the augmented data matrix is given by 
\begin{equation}\label{matrix_1}
\begin{pmatrix}0&1&x_{i1}&\hdots&x_{ip}\\0&2&x_{i1}&\hdots&x_{ip}\\0&3&x_{i1}&\hdots&x_{ip}\\ \vdots&\vdots&\vdots&&\vdots\\1&\tilde{T}_i&x_{i1}&\hdots&x_{ip}\end{pmatrix}.
\end{equation}
For an individual that is censored ($\Delta_i=0$) at time $\tilde{T}_i$ the augmented data matrix is given by 
\begin{equation}\label{matrix_0}
\begin{pmatrix}0&1&x_{i1}&\hdots&x_{ip}\\0&2&x_{i1}&\hdots&x_{ip}\\0&3&x_{i1}&\hdots&x_{ip}\\ \vdots&\vdots&\vdots&&\vdots\\0&\tilde{T}_i&x_{i1}&\hdots&x_{ip}\end{pmatrix}.
\end{equation}
The first column in the augumented data matrices corresponds to the binary responses $y_{i1},\hdots,y_{i\tilde{T}_i}$. The second column is the time interval running from $1$ to $\tilde{T}_i$. When fitting a model with fixed intercept parameters $\gamma_{0t}$ this column has necessarily to be coded as a nominal factor, e.g., via dummy variables. The remaining part of the data contains the covariates. When the covariates are constant over time, the values in each row of columns $3$ to $(p+2)$ are the same. This is also the case in the U.S.\@  unemployment data. Otherwise, when there are time-varying covariates, the observed time series are entered in the respective columns of the augmented data matrices. For each individual the augumented data matrix has $\tilde{T}_i$ rows, and the whole data matrix, which is obtained by ``glueing'' the individual augmented matrices together, has $\sum_{i=1}^{n}{\tilde{T}_i}$ rows. 

In \texttt{R} the augumented data matrix can be generated by applying the function \texttt{dataLong()} in the \texttt{R} package \textbf{discSurv} \citep{Welchov2015}. The general interface of the function is 
\begin{knitrout}\small
\definecolor{shadecolor}{rgb}{1, 1, 1}\color{fgcolor}\begin{kframe}
\begin{alltt}
\hlstd{> }\hlkwd{dataLong}\hlstd{(dataSet, timeColumn, censColumn,} \hlkwc{timeAsFactor}\hlstd{ = }\hlnum{TRUE}\hlstd{)}
\end{alltt}
\end{kframe}
\end{knitrout}
\noindent The function requires the original data of class \texttt{data.frame} in ``non-augumented'' short format (argument \texttt{dataSet}), the column name of the observed discrete event times (argument \texttt{timeColumn}) and the column name of the binary event indicator as defined in equation \eqref{delta} (argument \texttt{censColumn}). The variable required by \texttt{timeColumn} can either be numeric or coded as an ordinal or nominal factor. If \texttt{timeAsFactor = TRUE} the time column in the augumented data matrix will be returned as a nominal factor. The variable required by \texttt{censColumn} can either be a numerically coded 0/1 vector or a labeled factor variable. Note that \texttt{dataLong()} assumes that the covariates are constant over time. If this is not the case the function \texttt{dataLongTimeDep()} should be used instead to generate the augumented data matrix. 

The augumented data matrix returned by \texttt{dataLong()} contains the binary responses as defined in equation \eqref{responses} in the form of a numerically coded 0/1 vector named \texttt{y}. Further details on the output are given by the application in Section \ref{subsec:pre}. 

\section{Goodness-of-Fit Measures} \label{sec:gof}

In this tutorial we consider two diagnostic tools that are useful to investigate discrete hazard models in terms of their goodness of fit. First, one can generate a \textit{calibration plot}. The idea is to compare the estimated hazards $\hat{\lambda}(t|\xb_i), i=1,\hdots,n,\; t=1,\hdots,\tilde{T}_i$, of the model to the relative frequencies of observed events ($y_{it}=1$) in predefined subsets of the augmented set of observations. More specifically, one splits the data into subsets $D_k,\; k=1,\hdots,K$, defined by the percentiles of the estimated hazards. Common choices for $K$ are $K=10$ or $K=20$. Then the relative frequency of observed events (``empirical hazard'')  is calculated in each subset by 
\begin{equation}\label{relFreq}
\sum_{i,t: \hat{\lambda}(t|\xb_i) \in D_k} \frac{y_{it}}{|D_k|} \, ,
\end{equation}
where $|D_k|$ corresponds to the number of observations in subset $D_k$. If the fit of the model is satisfactory the empirical hazard measure in (\ref{relFreq}) should be close to the average of the estimated hazards in $D_k$ for all $k$. An example of a calibration plot is shown in Figure \ref{fig:goodness} in the application. 

Second, we consider \textit{martingale residuals}, which allow for assessing the importance of single covariates $x_j$. The idea of the martingale residuals is to compare for each individual the observed number of events with the expected number of events up to~$\tilde{T}_i$. Using the binary response variables $y_{i1},\hdots,y_{i\tilde{T_i}}$ the residuals are defined as 
\begin{equation}\label{martingale}
r_i=\sum_{t=1}^{\tilde{T}_i}(y_{it}-\hat{\lambda}(t|\xb_i)),\quad i=1,\hdots,n.
\end{equation}
For a well fitting model that includes all relevant predictors, the difference between $y_{it}$ and $\hat{\lambda}(t|\xb_i)$ should be ``random'' and therefore uncorrelated with the covariate values. To assess the importance of a covariate graphically one can plot the residuals against the covariate values. Martingale residuals can be computed by the function \texttt{martingaleResid()} contained in the \textbf{discSurv} package. An example is shown in Figure \ref{fig:goodness} in the application.

\section{Application: Duration of Unemployment} \label{sec:app}

In the following discrete hazard regression modeling is illustrated by means of a step-by-step analysis of the U.S.\@ unemployment data. Throughout this section we use the logistic link function, i.e., we consider the fitting of a proportional continuation ratio model.

\subsection{Preprocessing of the Data}\label{subsec:pre}

To fit a logistic discrete hazard model of the form \eqref{model_response}, the original data matrix first has to be transformed to an augmented data matrix, as described above. The data set \texttt{UnempDur}, which (after application of the pre-processing steps outlined in the Introduction) is a slightly modified version of the data frame available in the \texttt{R} package \texttt{Ecdat}, has the following form:   

\begin{knitrout}\small
\definecolor{shadecolor}{rgb}{1, 1, 1}\color{fgcolor}\begin{kframe}
\begin{alltt}
\hlstd{> }\hlkwd{head}\hlstd{(UnempDur)}
\end{alltt}
\begin{verbatim}
  spell age  ui reprate disrate logwage tenure status
1     5  41  no   0.179   0.045 6.89568      3      1
2    13  30 yes   0.520   0.130 5.28827      6      1
4     3  26 yes   0.448   0.112 5.97889      3      1
5     9  22 yes   0.320   0.080 6.31536      0      1
6    11  43 yes   0.187   0.047 6.85435      9      0
8     3  32  no   0.373   0.093 6.16121      0      1
\end{verbatim}
\end{kframe}
\end{knitrout}
\noindent The first column named \texttt{spell} is the observed time to re-employment of indiviudal $i$ and contains the values of $\tilde{T}_i,\; i=1,\hdots,n$. As mentioned above, these values correspond to the lengths of the spells (measured in two week intervals), whose distribution is displayed in Figure \ref{fig:data_time}. The last column named \texttt{status} indicates whether the exact event time of invidiual $i$ has been observed (\texttt{status} $=1$) or if the individual is subject to right censoring (\texttt{status} $=0$); it corresponds to the random variable $\Delta_i$ defined in equation \eqref{delta}. Summarizing the \texttt{status} column yields a censoring rate of~$0.391$:
\begin{knitrout}\small
\definecolor{shadecolor}{rgb}{1, 1, 1}\color{fgcolor}\begin{kframe}
\begin{alltt}
\hlstd{> }\hlkwd{table}\hlstd{(UnempDur}\hlopt{$}\hlstd{status)}\hlopt{/}\hlkwd{nrow}\hlstd{(UnempDur)}
\end{alltt}
\begin{verbatim}
        0         1 
0.3909657 0.6090343 
\end{verbatim}
\end{kframe}
\end{knitrout}
Columns two to five of the data frame \texttt{UnempDur} contain the explanatory variables described in Table \ref{tab:data_summary}. All covariates are constant over the time of the survey. 

\noindent When using \texttt{dataLong()} to obtain the augumented data matrix one has to pass the column names \texttt{spell} and \texttt{status} to the arguments \texttt{timeColumn} and \texttt{censColumn},
respectively: 
\begin{knitrout}\small
\definecolor{shadecolor}{rgb}{1, 1, 1}\color{fgcolor}\begin{kframe}
\begin{alltt}
\hlstd{> }\hlkwd{library}\hlstd{(discSurv)}
\hlstd{> }\hlstd{UnempDurLong} \hlkwb{<-} \hlkwd{dataLong}\hlstd{(UnempDur,} \hlkwc{timeColumn} \hlstd{=} \hlstr{"spell"}\hlstd{,}
\hlstd{+ }                         \hlkwc{censColumn} \hlstd{=} \hlstr{"status"}\hlstd{)}
\end{alltt}
\end{kframe}
\end{knitrout}
\noindent The augmented data matrix \texttt{UnempDurlong} has ten columns with the following names:  
\begin{knitrout}\small
\definecolor{shadecolor}{rgb}{1, 1, 1}\color{fgcolor}\begin{kframe}
\begin{alltt}
\hlstd{> }\hlkwd{names}\hlstd{(UnempDurLong)}
\end{alltt}
\begin{verbatim}
 [1] "obj"     "timeInt" "y"       "spell"   "age"     "ui"     
 [7] "reprate" "disrate" "logwage" "tenure"  "status" 
\end{verbatim}
\end{kframe}
\end{knitrout}
\noindent The new columns are \texttt{obj}, which is an identifier of the individuals, \texttt{timeInt}, which contains the discrete time values (i.e., the second column of the augmented matrices in \eqref{matrix_1} and \eqref{matrix_0}, stored as a nominal factor) and \texttt{y}, which contains the binary response variables $y_{i1},\hdots,y_{i\tilde{T_i}} \in \{0,1\}$. The head of the augmented data matrix is given by 
\begin{knitrout}\small
\definecolor{shadecolor}{rgb}{1, 1, 1}\color{fgcolor}\begin{kframe}
\begin{alltt}
\hlstd{> }\hlstd{UnempDurLong[UnempDurLong}\hlopt{$}\hlstd{obj}\hlopt{==}\hlnum{1}\hlstd{, ]}
\end{alltt}
\begin{verbatim}
    obj timeInt y spell age ui reprate disrate logwage tenure status
1     1       1 0     5  41 no   0.179   0.045 6.89568      3      1
1.1   1       2 0     5  41 no   0.179   0.045 6.89568      3      1
1.2   1       3 0     5  41 no   0.179   0.045 6.89568      3      1
1.3   1       4 0     5  41 no   0.179   0.045 6.89568      3      1
1.4   1       5 1     5  41 no   0.179   0.045 6.89568      3      1
\end{verbatim}
\end{kframe}
\end{knitrout}
\noindent showing that the first individual (\texttt{obj} $=1$) had an event after ten weeks (\texttt{spell} $=5$ and \texttt{status} $=1$).  Accordingly, the augumented data matrix for the first individual has five rows, where each row corresponds to one time interval (\texttt{timeInt} $ =1,\hdots,5$). The corresponding vector of responses is \texttt{y}$=(0,0,0,0,1)$. The values of the covariates remain constant over time and are therefore the same in each row.

As a second example, consider the augmented data matrix of the twelfth invidiual (\texttt{obj}$=12$). This individual is censored after 6 weeks (\texttt{spell}$=3$ and \texttt{status}$=0$), and hence the corresponding data matrix has three rows with response \texttt{y}$=(0,0,0)$:
\begin{knitrout}\small
\definecolor{shadecolor}{rgb}{1, 1, 1}\color{fgcolor}\begin{kframe}
\begin{alltt}
\hlstd{> }\hlstd{UnempDurLong[UnempDurLong}\hlopt{$}\hlstd{obj}\hlopt{==}\hlnum{12}\hlstd{, ]}
\end{alltt}
\begin{verbatim}
     obj timeInt y spell age  ui reprate disrate logwage tenure status
14    12       1 0     3  40 yes    0.52    0.13 4.95583      0      0
14.1  12       2 0     3  40 yes    0.52    0.13 4.95583      0      0
14.2  12       3 0     3  40 yes    0.52    0.13 4.95583      0      0
\end{verbatim}
\end{kframe}
\end{knitrout}

\subsection{Regression Modeling} \label{subsec:models}

A parametric proportional continuation ratio model with a linear predictor is estimated in \texttt{R} by passing the augumented data matrix \texttt{UnempDurLong} to \texttt{glm()} with the usual specifications:
\begin{knitrout}\small
\definecolor{shadecolor}{rgb}{1, 1, 1}\color{fgcolor}\begin{kframe}
\begin{alltt}
\hlstd{> }\hlstd{model1} \hlkwb{<-} \hlkwd{glm}\hlstd{(}\hlkwc{formula} \hlstd{= y} \hlopt{~} \hlstd{timeInt} \hlopt{-} \hlnum{1} \hlopt{+}
\hlstd{+ }              \hlstd{age} \hlopt{+} \hlstd{reprate}  \hlopt{+} \hlstd{disrate} \hlopt{+} \hlstd{logwage} \hlopt{+} \hlstd{tenure} \hlopt{+} \hlstd{ui,}
\hlstd{+ }              \hlkwc{data} \hlstd{= UnempDurLong,} \hlkwc{family} \hlstd{=} \hlkwd{binomial}\hlstd{(}\hlkwc{link} \hlstd{=} \hlstr{"logit"}\hlstd{))}
\end{alltt}
\end{kframe}
\end{knitrout}
\noindent The left hand side of the \texttt{formula} argument contains the binary response vector~\texttt{y}. In addition to the names of the six covariates, the right hand side contains the discrete time variable as a nominal factor without intercept (\texttt{timeInt - 1}). The \texttt{family} argument \texttt{binomial(link = "logit")} is the same as for ``usual'' logistic regression models with binary outcome.

The more complex model with smooth baseline hazard (equation \eqref{eta_smooth_base}) is estimated by use of the \texttt{R} package \textbf{mgcv} \citep{mgcv2017}. A detailed introduction to the estimation procedures is found in \citet{Wood2011}. The corresponding function \texttt{gam()} essentialy has the same interface as \texttt{glm()}:  
\begin{knitrout}\small
\definecolor{shadecolor}{rgb}{1, 1, 1}\color{fgcolor}\begin{kframe}
\begin{alltt}
\hlstd{> }\hlkwd{library}\hlstd{(}\hlstr{"mgcv"}\hlstd{)}
\hlstd{> }\hlstd{UnempDurLong}\hlopt{$}\hlstd{timeIntNum} \hlkwb{<-} \hlkwd{as.numeric}\hlstd{(UnempDurLong}\hlopt{$}\hlstd{timeInt)}
\hlstd{> }\hlstd{model2} \hlkwb{<-} \hlkwd{gam}\hlstd{(}\hlkwc{formula} \hlstd{= y} \hlopt{~} \hlkwd{s}\hlstd{(timeIntNum,} \hlkwc{bs} \hlstd{=} \hlstr{"ps"}\hlstd{,} \hlkwc{k} \hlstd{=} \hlnum{5}\hlstd{,} \hlkwc{m} \hlstd{=} \hlnum{2}\hlstd{)} \hlopt{+}
\hlstd{+ }              \hlstd{age} \hlopt{+} \hlstd{reprate}  \hlopt{+} \hlstd{disrate} \hlopt{+} \hlstd{logwage} \hlopt{+} \hlstd{tenure} \hlopt{+} \hlstd{ui,}
\hlstd{+ }              \hlkwc{data} \hlstd{= UnempDurLong,} \hlkwc{family} \hlstd{=} \hlkwd{binomial}\hlstd{(}\hlkwc{link} \hlstd{=} \hlstr{"logit"}\hlstd{))}
\end{alltt}
\end{kframe}
\end{knitrout}
\noindent Before passing the nominal factor \texttt{timeInt} to \texttt{gam()}, it has to be transformed to a continuous variable (\texttt{timeIntNum}). Then a smooth baseline hazard is specified by use of the function \texttt{s()} on the right hand side of the model formula. Required arguments are the type of spline smoother \texttt{bs}, the dimension of the basis \texttt{k}, which determines the number of basis functions, and the order of the penalty \texttt{m}.  Here we use $P$-splines with cubic basis functions (\texttt{bs = "ps"}) and a second order difference penalty (\texttt{m} $=2$). The chosen dimension (\texttt{k} $=5$) results in nine cubic basis functions. Based on these specifications the optimal smoothing parameter $\delta$, see equation \eqref{pen_ll}, is computed by generalized cross-validation, see \citet{Wood2006}. Its estimate is stored in the argument \texttt{sp}. In case of the U.S.\@ unemployment data, \texttt{sp} is estimated as:
\begin{knitrout}\small
\definecolor{shadecolor}{rgb}{1, 1, 1}\color{fgcolor}\begin{kframe}
\begin{alltt}
\hlstd{> }\hlstd{model2}\hlopt{$}\hlstd{sp}
\end{alltt}
\begin{verbatim}
s(timeIntNum) 
   0.08562284 
\end{verbatim}
\end{kframe}
\end{knitrout}
\noindent Generally, the {\bf mgcv} package implements a large variety of alternative spline estimators and methods for smoothing parameter optimization. In principle, all of these methods may used for discrete hazard modeling, in the same way as they would be used in logistic regression (or, more generally, in additive models with a binary response).

\begin{figure}[!t]
\centering
\includegraphics[width=0.5\textwidth]{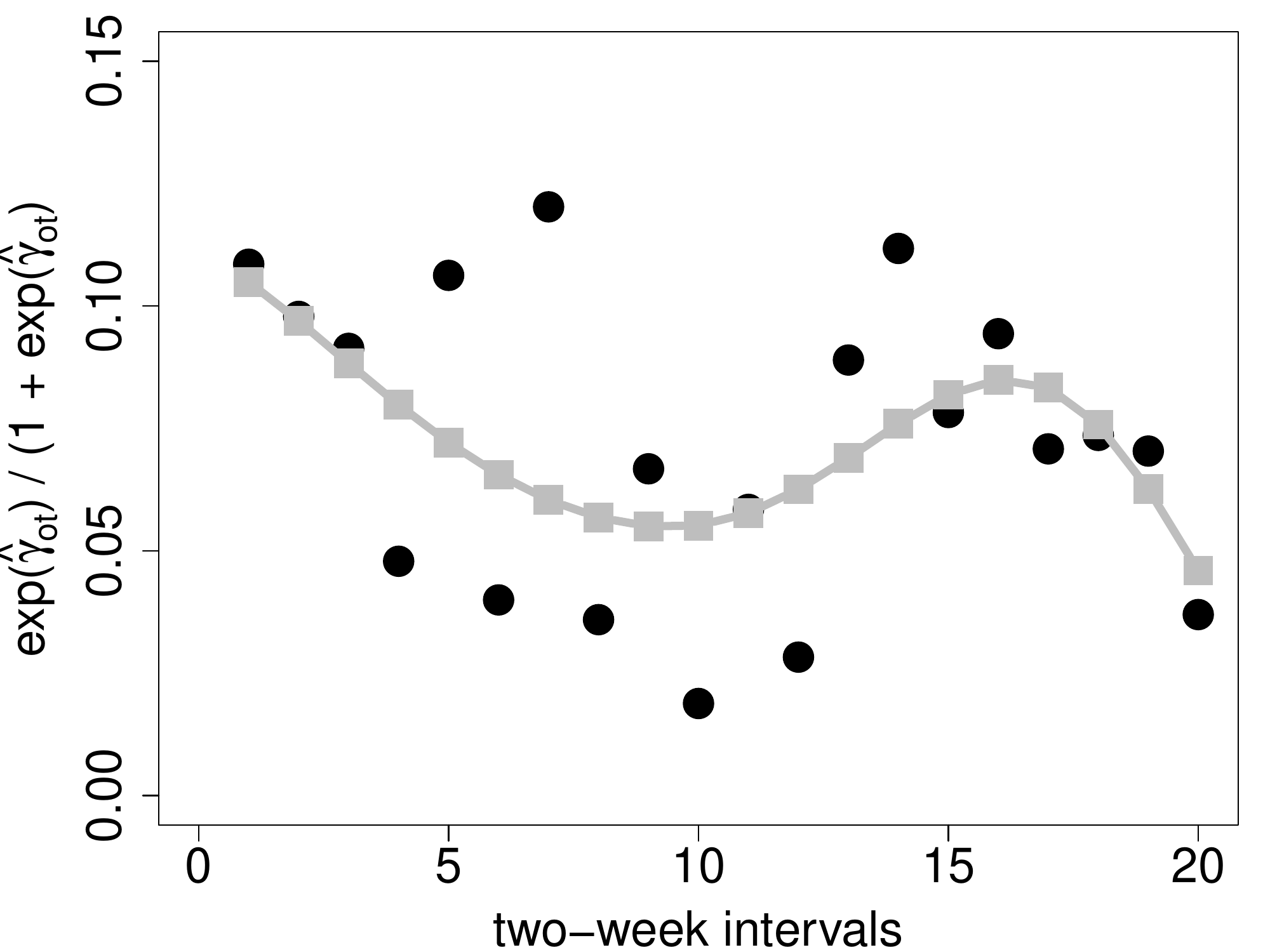}
\caption{Analysis of the U.S.\@ unemployment data. The figure shows the estimated discrete baseline hazard of \texttt{model1} (black dots) and the smooth baseline hazard of \texttt{model2} (grey squares).}
\label{fig:baseline_hazard}
\end{figure}

The estimated baseline hazards of \texttt{model1} and \texttt{model2} are shown in Figure \ref{fig:baseline_hazard}. The discrete baseline hazard obtained for \texttt{model1} is visualized by black dots. From these estimates a reasonable interpretation is hard to derive. On the other hand, the smooth baseline hazard obtained for \texttt{model2} visualized by grey squares is more meaningful. It is seen that the conditional probability of re-employment decreases until week 20 and subsequently increases up to week 32 before it diminishes again. The reason for this might be that in many U.S. states workers are eligible for up to 26 weeks of benefits from the state-funded unemployment compensation program.

\begin{table}[!t]
\caption{Analysis of the U.S.\@ unemployment data. The table contains coefficient estimates (coef), estimated standard errors (se) and $p$-values of the covariate effects obtained for \texttt{model1} and \texttt{model2} (bh = baseline hazard).}
\begin{center}
\begin{tabularsmall}{p{1.5cm}rrrp{0.25cm}rrr}
\toprule
&\multicolumn{3}{c}{\bf{model1 (discrete bh)}}&&\multicolumn{3}{c}{\bf{model2 (smooth bh)}}\\
&coef&se&$p$-value&&coef&se&$p$-value\\ 
\midrule
age  & -0.012 & 0.003 & 0.000 && -0.012 & 0.003 & 0.000\\ 
reprate  & 0.285 & 0.342 & 0.406 && 0.301 & 0.338 & 0.373\\ 
disrate  & -0.764 & 0.383& 0.046 && -0.755 & 0.379 & 0.047\\ 
logwage   & 0.231 & 0.072 & 0.001 && 0.236 & 0.071 & 0.001\\ 
tenure  & -0.005 & 0.005  & 0.280 && -0.006 & 0.005 & 0.266\\ 
ui  & -1.151 & 0.052  & 0.000 && -1.175 & 0.051 & 0.000\\ 
\bottomrule
\end{tabularsmall}
\label{tab:est_model12}
\end{center}
\end{table}

Table \ref{tab:est_model12} shows the estimates of the coefficients $\gammab$, the corresponding estimated standard errors, and the $p$-values of the covariate effects obtained for \texttt{model1} and \texttt{model2}. Apart from the eligible replacement rate (\texttt{reprate}) and the tenure in the lost job (\texttt{tenure}), the covariates are significantly associated with the time to re-employment. According to the signs of the estimates of both models, the chance of getting re-employed decreases with increasing values of age, increasing disregard rate and with the filing of an unemployment claim. On the other hand, the higher the earnings in the lost job the better the chance of re-employment. Table \ref{tab:est_model12} also shows that the differences in coefficient estimates between the two models are small. By use of equation \eqref{model_continuation} the effects can be interpretated in an easy way: For example, let us compare citizens who submitted an unemployment insurance claim (\texttt{ui} $=1$) to those who did not (\texttt{ui} $=0$). Based on the estimate of \texttt{model1} ($\hat{\gamma}_{ui}=-1.151$), one obtains that the probability of re-employment at time $t$, compared to the probability of re-employment later than $t$, decreases for citizens who submitted an unemployment insurance claim by the factor $\exp(\hat{\gamma}_{ui})=0.316$. The chance of re-emplyoment is therefore much smaller in this group. One might speculate that due to benefits from the state-funded unemployment the motivation to search for a new job is lower.

\begin{figure}[!t]
\centering
\includegraphics[width=0.49\textwidth]{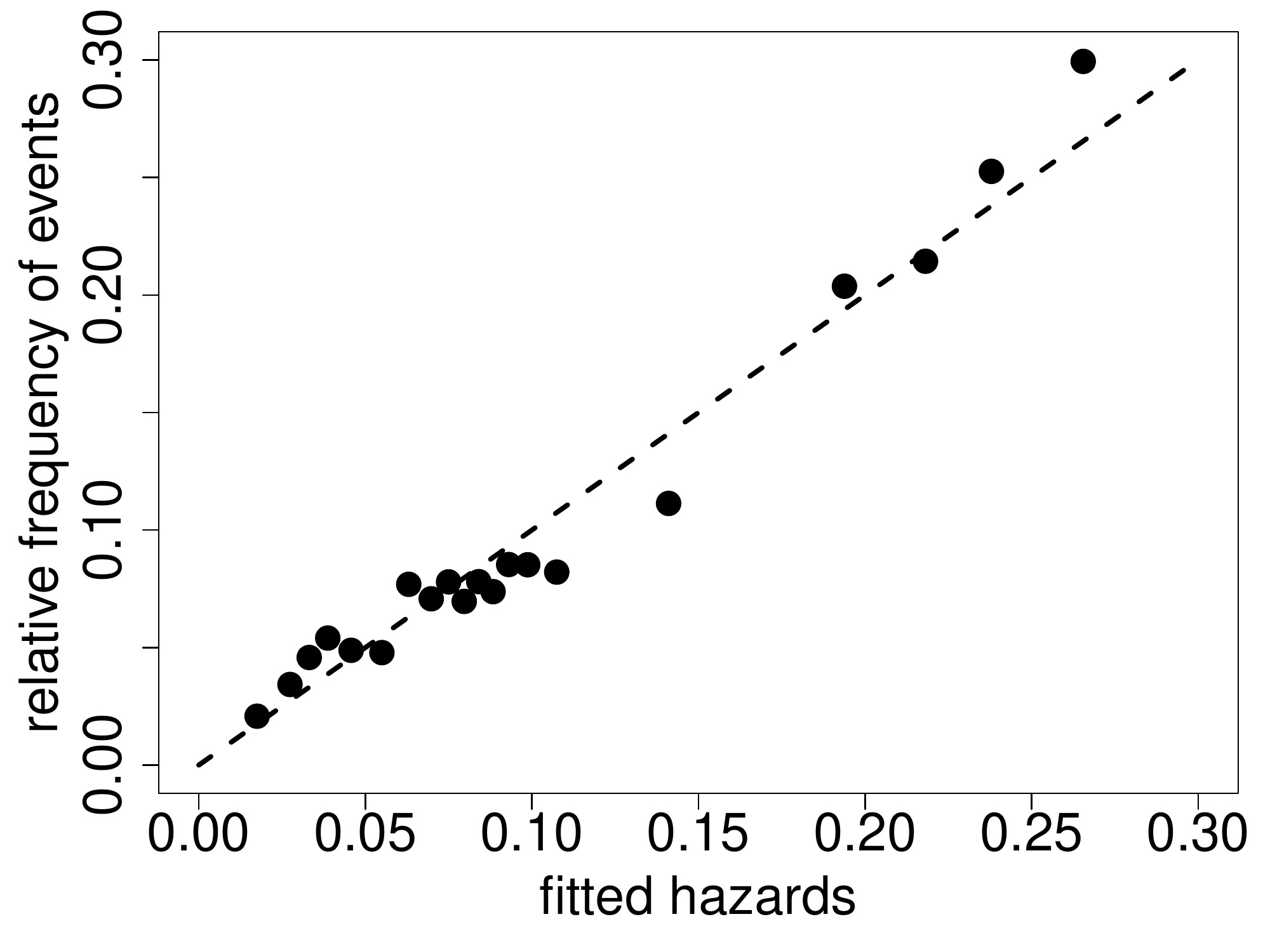}
\includegraphics[width=0.49\textwidth]{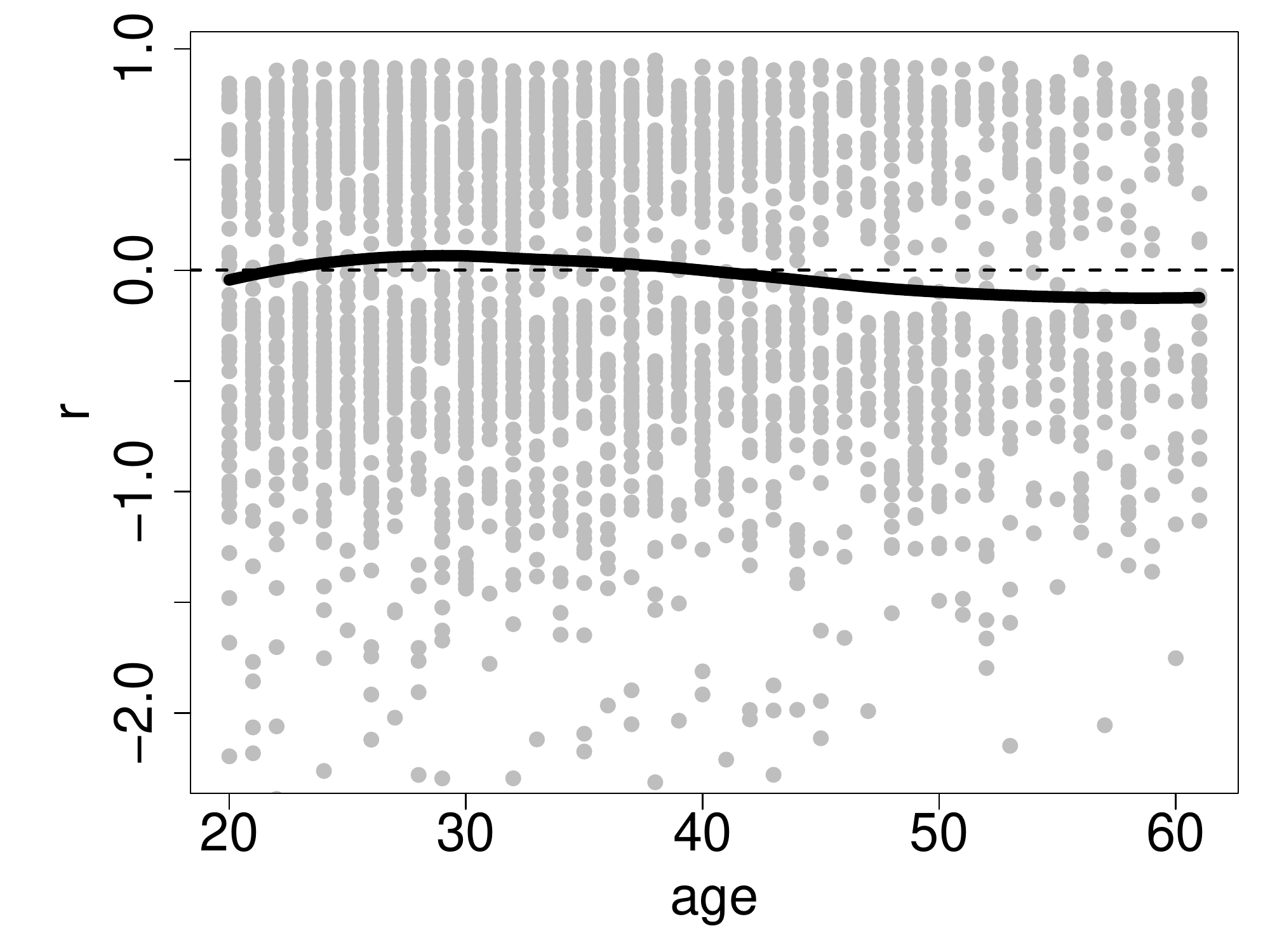}
\caption{Analysis of the U.S.\@ unemployment data. The two panels show the calibration plot (left) for \texttt{model1} and the martingale residuals against the values of age (right) obtained for \texttt{model1} without age. The trend line was obtained by local polynomial regression.}
\label{fig:goodness}
\end{figure}

The goodness-of-fit measures for \texttt{model1} are presented in Figure \ref{fig:goodness}. The left panel shows the calibration plot (average fitted hazards against the relative frequencies of events). It is seen that the values do not deviate too much from the 45 degree line, indicating an acceptable model fit. The right panel shows the martingale residuals defined in \eqref{martingale}, against the values of the covariate \texttt{age} for \texttt{model1} without age. The black line corresponds to the estimated trend obtained by a local polynomial regression using the \texttt{R} function \texttt{loess()}. The functional form of the trend line (compared to the zero line) shows a non-linear effect on the martingale residuals. This indicates that the covariate \texttt{age} is an influential variable with a non-linear effect on the response.

\begin{figure}[!t]
\centering
\includegraphics[width=0.5\textwidth]{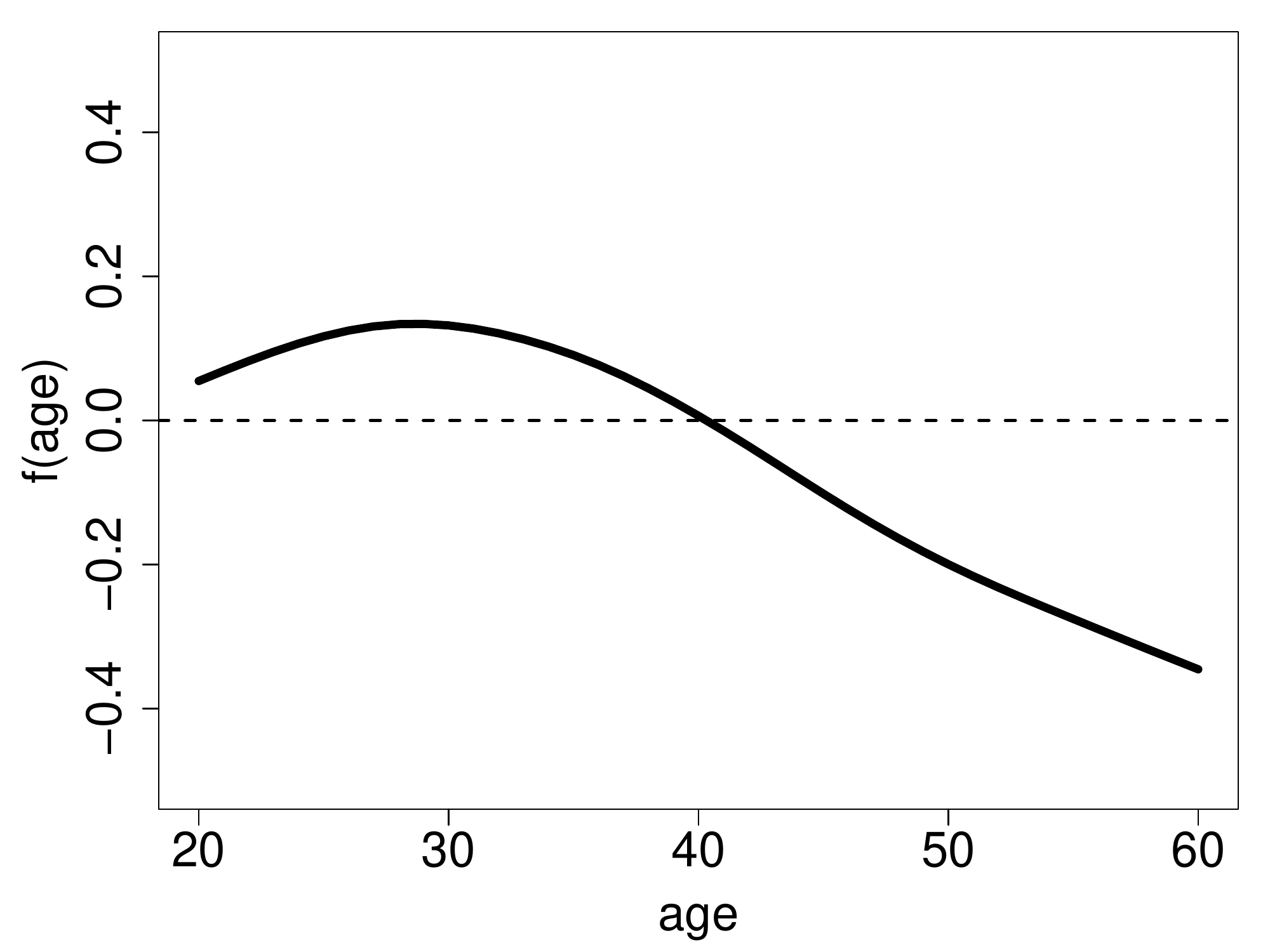}
\caption{Analysis of the U.S.\@ unemployment data. The plot shows the estimated $P$-spline function for the covariate age in \texttt{model3}.}
\label{fig:smooth_age}
\end{figure}

Therefore, as a possible extension, we consider a model where the baseline hazard as well as the covariate age are both modeled as smooth $P$-spline functions:
\begin{knitrout}\small
\definecolor{shadecolor}{rgb}{1, 1, 1}\color{fgcolor}\begin{kframe}
\begin{alltt}
\hlstd{> }\hlstd{model3} \hlkwb{<-} \hlkwd{gam}\hlstd{(}\hlkwc{formula} \hlstd{= y} \hlopt{~} \hlkwd{s}\hlstd{(timeIntNum,} \hlkwc{bs} \hlstd{=} \hlstr{"ps"}\hlstd{,} \hlkwc{k} \hlstd{=} \hlnum{5}\hlstd{,} \hlkwc{m} \hlstd{=} \hlnum{2}\hlstd{)} \hlopt{+}
\hlstd{+ }              \hlkwd{s}\hlstd{(age,} \hlkwc{bs} \hlstd{=} \hlstr{"ps"}\hlstd{,} \hlkwc{k} \hlstd{=} \hlnum{25}\hlstd{,} \hlkwc{m} \hlstd{=} \hlnum{2}\hlstd{)} \hlopt{+}
\hlstd{+ }              \hlstd{reprate}  \hlopt{+} \hlstd{disrate} \hlopt{+} \hlstd{logwage} \hlopt{+} \hlstd{tenure} \hlopt{+} \hlstd{ui,}
\hlstd{+ }              \hlkwc{data} \hlstd{= UnempDurLong,} \hlkwc{family} \hlstd{=} \hlkwd{binomial}\hlstd{(}\hlkwc{link} \hlstd{=} \hlstr{"logit"}\hlstd{))}
\end{alltt}
\end{kframe}
\end{knitrout}
\noindent As seen from the \texttt{R} code, the estimation of the smooth function of covariate age in \texttt{model3} is based on 29 cubic basis functions (dimension \texttt{k} $=25$). The estimated penalty parameter $\hat{\delta}$ for age, stored in \texttt{sp}, is:
\begin{knitrout}\small
\definecolor{shadecolor}{rgb}{1, 1, 1}\color{fgcolor}\begin{kframe}
\begin{alltt}
\hlstd{> }\hlstd{model3}\hlopt{$}\hlstd{sp[}\hlstr{"s(age)"}\hlstd{]}
\end{alltt}
\begin{verbatim}
 s(age) 
56860.2 
\end{verbatim}
\end{kframe}
\end{knitrout}
\noindent From the resulting function shown in Figure \ref{fig:smooth_age} it is seen that the association between the time to re-employment and age is definitely not linear. Its form is very similar to the loess trend shown in Figure \ref{fig:goodness}. The value on the $y$-axis of the figure corresponds to the contribution of age to the predictor $\eta_{it}$ of the model. The chance of re-employment has a peak between 20 years and 30 years and subsequently decreases.

\subsection{Tree-Based Modeling}\label{subsec:tree}

Finally, we fit a recursive partitioning model of the form \eqref{eta_tree}. Again we consider a procedure that is based on the augmented data matrix with binary outcomes $y_{i1},\hdots,y_{i\tilde{T_i}}$.

When growing trees one has to take two main decisions: Firstly, one has to choose an appropriate  criterion for performing the splits. Criteria that have already been used in the early days of tree construction are impurity measures. For discrete surival trees a natural measure of node impurity is the {\em Brier score}, which evaluates the average squared difference between the binary outcome values $y_{it}$ and the respective hazard estimate $\hat{\lambda} (t|\xb_i )$ in each node, see \citet{Schmid2016}. It can be shown that using the Brier score is equivalent to the traditional Gini impurity measure. For a single node $m$ the Gini impurity is given by 
\begin{equation}
G_m=2\, \pi_m \, (1-\pi_m),
\end{equation}
where $\pi_m$ is the proportion of ones in node $m$, see \citet{Breiman1996}. This equivalence implies that the traditional CART algorithm based on the Gini criterion can be used for the construction of the tree. The latter is done by using the function \texttt{rpart()} of the eponymous \texttt{R} package \textbf{rpart} \citep{rpart2015}.

Secondly, one has to determine the optimal size of the tree. For the discrete survival tree an appropriate tuning parameter controlling tree size is the minimal number of observations in each terminal node (``minimal node size''). Optimizing this number avoids overfitting, as the number of terminal nodes is prevented from becoming too large and, at the same time, the node sizes are prevented from becoming too small. Accordingly, splitting is stopped when further splitting in any of the current nodes would result in an additional node containing less observations than the minimal node size. Given a sequence of tree estimates depending on the minimal node size, the optimal tree (i.e., the tree with ``optimal'' minimal node size) is determined by either minimization of an information criterion (such as AIC or BIC, see below) or maximization of the predictive log-likelihood. The latter strategy means to repeatedly draw subsamples from the original non-augmented data (for example, by cross-validation, bootstrapping or subsampling without replacement), and to calculate the log-likelihood for the omitted observations. One determines the optimal tree as the one for which the predictive log-likelihood (averaged across the subsamples) becomes maximal. The \texttt{R} function \texttt{survivalTree()} automatically generates the augumented data matrix by \texttt{dataLong()}, estimates the discrete survival tree by \texttt{rpart()} und returns the optimal one according to the specified performance criterion. The function is part of the electronic supplement of this article. 

Once the optimal minimal node size has been determined, the estimate of $\lambda(t|\xb_i)$ is given  by the relative frequency of events (proportion of ones) in each node, possibly after applying some sort of correction procedure like the Laplace correction (see below).

To fit a tree model to the U.S.\@ unemployment data, we call the \texttt{survivalTree} function using the following arguments: 
\begin{knitrout}\small
\definecolor{shadecolor}{rgb}{1, 1, 1}\color{fgcolor}\begin{kframe}
\begin{alltt}
\hlstd{> }\hlkwd{source}\hlstd{(}\hlstr{"survivalTree.R"}\hlstd{)}
\hlstd{> }\hlstd{model4} \hlkwb{<-} \hlkwd{survivalTree}\hlstd{(}\hlkwc{formula} \hlstd{= y} \hlopt{~} \hlstd{timeInt} \hlopt{+} \hlstd{age} \hlopt{+}
\hlstd{+ }                       \hlstd{reprate} \hlopt{+} \hlstd{disrate} \hlopt{+} \hlstd{logwage} \hlopt{+} \hlstd{tenure} \hlopt{+} \hlstd{ui,}
\hlstd{+ }                       \hlkwc{data} \hlstd{= UnempDur,} \hlkwc{tuning} \hlstd{=} \hlstr{"BIC"}\hlstd{,}
\hlstd{+ }                       \hlkwc{timeColumn} \hlstd{=} \hlstr{"spell"}\hlstd{,} \hlkwc{censColumn} \hlstd{=} \hlstr{"status"}\hlstd{,}
\hlstd{+ }                       \hlkwc{minimal_ns} \hlstd{=} \hlkwd{seq}\hlstd{(}\hlnum{100}\hlstd{,} \hlnum{1500}\hlstd{,} \hlkwc{by}\hlstd{ = }\hlnum{10}\hlstd{),}
\hlstd{+ }                       \hlkwc{trace} \hlstd{=} \hlnum{TRUE}\hlstd{)}
\end{alltt}
\end{kframe}
\end{knitrout}
\noindent The formula required for the tree model is analogous to the one specified for a model with linear predictor. Note that internally the time variable \texttt{timeInt} is coded as a numeric vector. This is in analogy to \texttt{model2} and \texttt{model3} with smooth baseline hazard. The original data frame \texttt{UnempDur} (in non-augmented format) is passed to the \texttt{data} argument. In addition one has to specify the \texttt{timeColumn} and \texttt{CensColumn} arguments used in \texttt{dataLong()}. The performance criterion is specified by the argument \texttt{tuning}. For tuning we use the Bayesian information criterion (BIC) defined by
\[
\mathrm{BIC} := -2\,l + \log(\tilde{n})\,n_s\,,
\]
where $l$ is the log-likelihood \eqref{logL}, $\tilde{n}$ is the number of rows of the augumented data matrix, and $n_s$ denotes the number of splits as a measure of the complexity of the tree. Other possible arguments for tuning are \texttt{"AIC"} (Akaike's information criterion) and \texttt{"ll"} (predictive log-likelihood method). When using \texttt{"ll"}, \texttt{survivalTree()} performs a five-fold cross-validation based on subsamples without replacement stratified by \texttt{spell}. The \texttt{survivalTree} function searches for the best model among the sequence of models with minimal node sizes  \texttt{minimal\_ns}. If \texttt{minimal\_ns} is not specified, the sequence of minimal node sizes is set to $1,\hdots, \lfloor{\tilde{n}/2}\rfloor$. 

\begin{figure}[!t]
\centering
\includegraphics[width=0.5\textwidth]{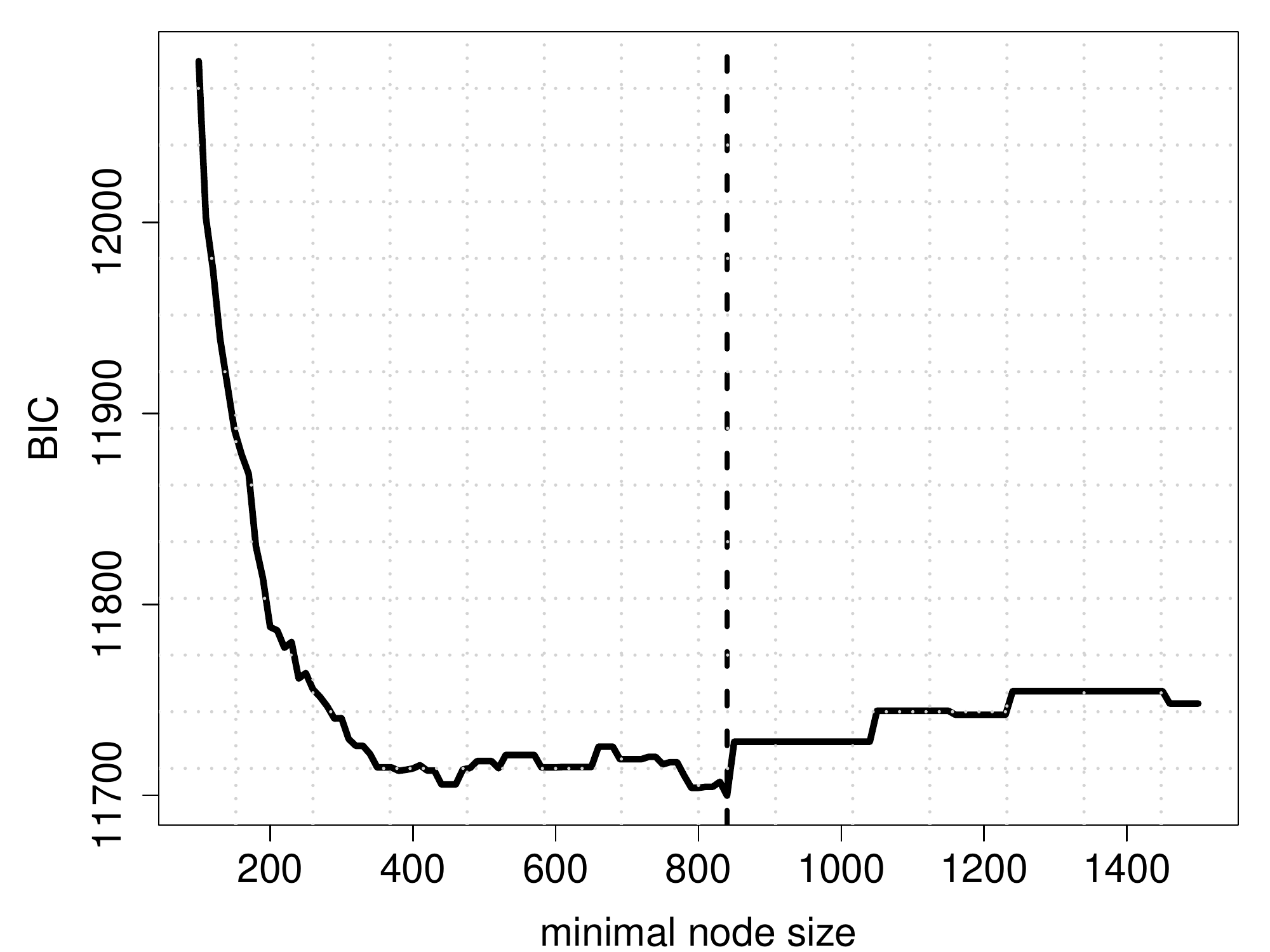}
\caption{Analysis of the U.S.\@ unemployment data. The plot shows the BIC values for the sequence of survival trees that was obtained by fitting \texttt{model4} with minimal node sizes ranging from 100 to 1500. The minimal BIC value (obtained for node size 840) is marked by the vertical dashed line.}
\label{fig:BICs}
\end{figure}

\begin{figure}[!t]
\centering
\includegraphics[trim= 0mm 50mm 0mm 0mm, width=1\textwidth]{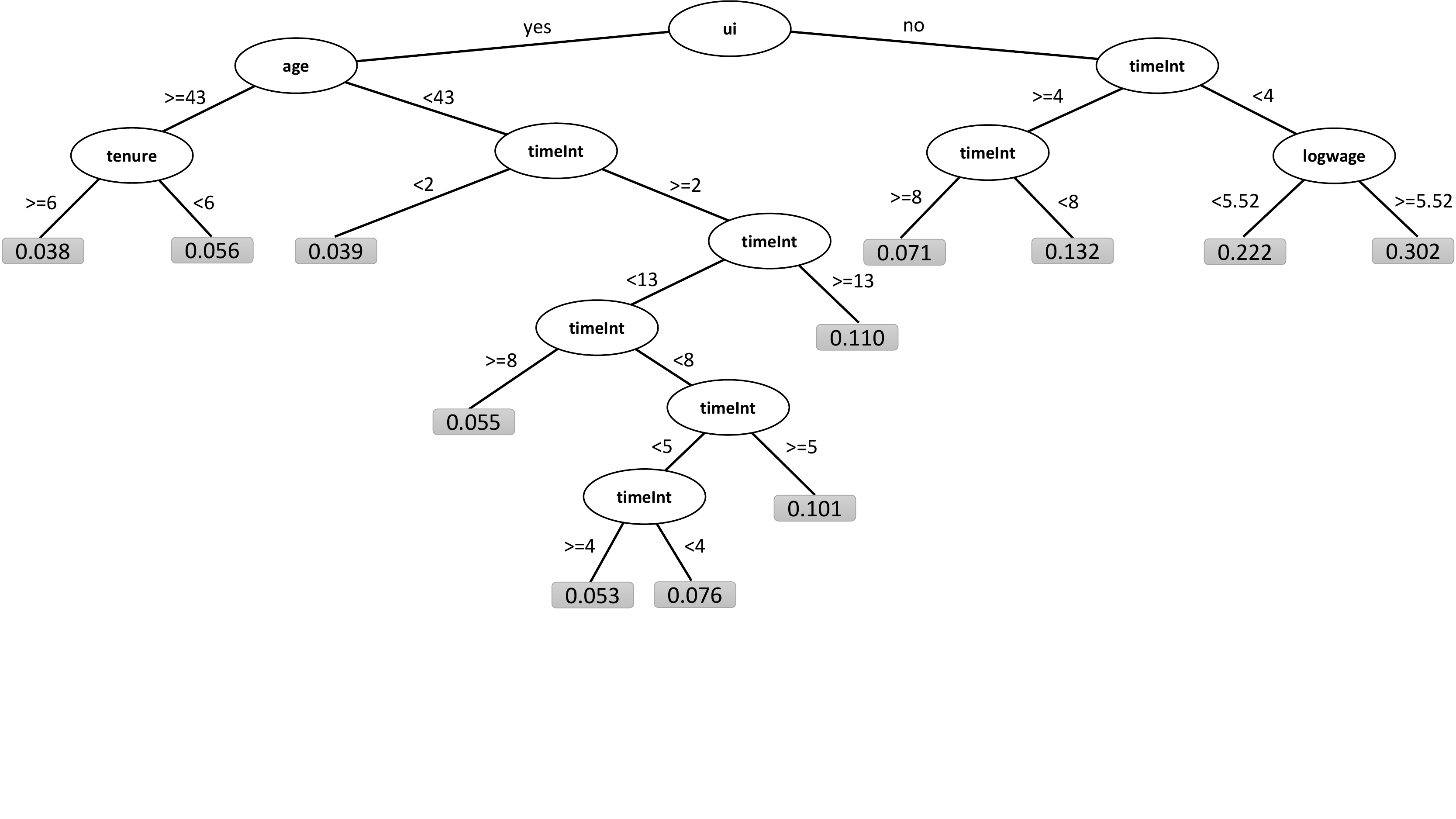}
\caption{Analysis of the U.S.\@ unemployment data. The graph visualizes the survival tree obtained from fitting \texttt{model4} with BIC-optimal minimal node size 840. The numbers at the terminal nodes refer to the estimated hazards. All estimated hazards were additionally post-processed by application of the {\em Laplace correction}, which was suggested by \citet{Ferri2003} to correct for estimates near the boundaries 0 and 1 in nodes with very few observations. The Laplace correction is automatically performed by \texttt{survivalTree()}.} 
\label{fig:tree2}
\end{figure}

The BICs obtained for \texttt{model4} with minimal node sizes $100,\hdots,1500$ (in steps of 10) are shown in Figure \ref{fig:BICs}. If an increase of the minimal node size does not change the number of splits and therefore does not influence the resulting tree, the BIC remains the same. This is the case, for example, between minimal node sizes $900$ and $1000$. According to the BIC, the optimal tree model has minimal node size 840, marked by the dashed line in Figure \ref{fig:BICs}. This results in a tree with eleven splits or twelve terminal nodes. The estimated tree is shown in Figure \ref{fig:tree2}. 

The most important covariate, which was chosen in the first split of the tree, is \texttt{ui}. As already derived from the parametric models, the submission of an unemployment insurance claim (\texttt{ui} = "yes") has a negative effect on the ``chance'' of re-employment. Within the group of citizens who submitted an insurance claim, the chance is lowest for citizens aged $43$ years or older and with a tenure in the lost job of at least 6 years (leftmost node in Figure~\ref{fig:tree2}). For citizens younger than $43$ years all further splits are performed with regard to the discrete time variable (\texttt{timeInt}). This confirms the results from Figure \ref{fig:baseline_hazard} in that the chance of re-employment is highly time-dependent. With a hazard estimate of $0.110$ after 26 weeks (\texttt{timeInt} $>= 13$) of unemployment in this group, the tree estimate also reflects the marked increase in the chance of re-employment already seen in Figure \ref{fig:baseline_hazard} after 20 weeks. The best opportunities of re-employment are observed for citizens without an unemployment insurance claim, within the first six weeks (\texttt{timeInt} $<4$) of unemployment, and with log weekly earnings of at least $5.52 \$$ (rightmost node in Figure~\ref{fig:tree2}). For this subgroup the estimated hazard rate is $0.302$. The two covariates \texttt{reprate} and \texttt{disrate} were not selected in any of the splits and are therefore exluded from the model. This is in contrast to \texttt{model1} and \texttt{model2} (see Table \ref{tab:est_model12}), where \texttt{disrate} showed a significant effect on the hazard.

\section{Concluding Remarks}\label{sec:remarks}

In this tutorial we have described a basic set of tools to fit semiparametric regression models with a discrete time-to-event outcome. All presented models are very general, in that they are applicable to any type of censored discrete response, regardless of whether the data-generating process is defined by an intrinsically discrete process or by the rounding/grouping of continuous event times. Furthermore, the presented methods are applicable in basically any field of research, as for example, in the social sciences, biostatistics, epidemiology, and many more. The U.S.\@ unemployment data considered in this article is therefore only one of many possible examples. Further applications are presented in \citet{Tutz2016}.

It is important to realize that all models considered in this tutorial can be fitted easily by use of standard software for binary regression modeling. The most important functions in \texttt{R} are \texttt{glm()}, \texttt{gam()} (of the \textbf{mgcv} package), and \texttt{rpart()} (of the eponymous package). In addition to the $P$-spline and CART methodologies considered here, many other options for semiparametric discrete time-to-event modeling exist in \texttt{R}. For example, \textbf{mgcv} provides a variety of alternative spline modeling tools such as cardinal splines and smoothing splines, which can be used for discrete hazard modeling by specifying the \texttt{bs} argument in \texttt{gam()} accordingly. Similarly, there is an alternative tree modeling approach developed by \citet{Bou2009} that operates directly on the non-augmented time-to-event data. This procedure is implemented in the \texttt{R} package \textbf{DStree} \citep{DStree2014}.

The basic functionalities required for applying the aforementioned software packages are all implemented in the \textbf{discSurv} package. Next to the functions used in this tutorial, \textbf{discSurv} provides additional functions to calculate, for example, measures for model evaluation like the concordance index \citep{Schmid2017}, and alternative tools for residual analysis.

We finally note that there exist a number of additional modeling options that are beyond the scope of this tutorial. These include, among many others, (i) regularized estimation via penalized optimization of the log-likelihood, which is useful for variable selection in higher-dimensional settings, (ii) random-effects and finite mixture modeling, which account for unobserved heterogeneity in the data, and (iii) competing-risks models, which extend the models considered in this tutorial by allowing for more than one target event. For details on further methodology, including semiparametric extensions, see \citet{Tutz2016}. 

\section*{Acknowledgements}

The work of MS was supported by the German Research Foundation (DFG), grant SCHM 2966/1-2.

\bibliography{literatur}

\end{document}